\documentclass[conference]{IEEEtran}
\IEEEoverridecommandlockouts
\usepackage{cite}
\usepackage{amsmath,amssymb,amsfonts}
\usepackage{algorithmic}
\usepackage{graphicx}
\usepackage{textcomp}
\usepackage{xcolor}
\def\BibTeX{{\rm B\kern-.05em{\sc i\kern-.025em b}\kern-.08em
    T\kern-.1667em\lower.7ex\hbox{E}\kern-.125emX}}

\newcommand{\name}{HEAL}
\newcommand{\newD}{Missed\_updates}
\newcommand{\LDP}{Partially\_persisted}

\definecolor{ao}{rgb}{0.0, 0.5, 0.0}

\usepackage{tikz}
\definecolor{lightmintbg}{rgb}{1,1,1}
\newcommand*\circleint[1]{\tikz[baseline=(char.base)]{
            \node[shape=circle,draw,inner sep=1pt, fill=lightmintbg] (char) {\footnotesize #1};}}

\usepackage{soul}
\usepackage{color} 
\usepackage{graphicx}
\usepackage{tabularx}
\usepackage{colortbl}
\usepackage{tabularx}
\usepackage{multirow}
\usepackage{pifont}
\usepackage{makecell} 
\usepackage[hidelinks]{hyperref} 
\usepackage{cleveref}
\usepackage{color,soul}
\usepackage{caption}
\usepackage{multicol}
\usepackage{comment}
\usepackage{csquotes}
\usepackage{subcaption}
\crefformat{section}{\S#2#1#3} 
\crefformat{subsection}{\S#2#1#3}
\crefformat{subsubsection}{\S#2#1#3}

\newcommand{\mytitle}{\name: Online Incremental Recovery for Leaderless Distributed Systems Across Persistency Models\vspace{-5mm}}

\begin{document}

\title{\huge\mytitle}

\author{
\IEEEauthorblockN{
Antonis Psistakis\IEEEauthorrefmark{1},
Burak Ocalan\IEEEauthorrefmark{1},
Fabien Chaix\IEEEauthorrefmark{2},
Ramnatthan Alagappan\IEEEauthorrefmark{1},
Josep Torrellas\IEEEauthorrefmark{1}
}
\IEEEauthorblockA{\IEEEauthorrefmark{1}University of Illinois Urbana-Champaign, USA}
\IEEEauthorblockA{\IEEEauthorrefmark{2}Foundation for Research and Technology-Hellas, Greece}
}
\vspace{-7mm}

\maketitle



\thispagestyle{plain} 
\pagestyle{plain}  


\begin{abstract}
Ensuring resilience in distributed systems
has become an acute concern.
In today’s environment,
it is crucial to develop light-weight mechanisms that recover
a distributed system from faults quickly and with only a small impact on the live-system
throughput.
To address this need, this paper proposes a new low-overhead, general
recovery scheme for modern {\em non-transactional   leaderless}  distributed
systems. We call our scheme {\em \name{}}. On a node failure, \name{} performs an 
optimized  online incremental  recovery. 
This paper presents \name{}’s
algorithms for settings with
Linearizable consistency and different memory persistency models. 
We implement \name{} on a 6-node Intel cluster. 
Our experiments running TAOBench workloads show that \name{} is very effective. 
\name{} recovers the cluster in 120
milliseconds on average, while reducing the throughput of the running
workload by an average of 8.7\%. 
In contrast, a conventional recovery scheme for   leaderless
systems needs 360 seconds to recover,
reducing the throughput of the system by 
16.2\%. 
Finally, compared to an incremental
recovery scheme for a state-of-the-art
{\em leader-based} system, \name{} reduces the average recovery latency by 20.7$\times$
and the throughput degradation by 62.4\%.

\end{abstract}

\section{Introduction}\label{sec:intro}

Distributed storage applications such as key value stores  
and databases \cite{tao-atc13, dynamo-sosp07, bigtable-acm08} are central
to today's cloud infrastructure. To provide 
high performance and availability, these applications replicate data
across multiple nodes~\cite{farm-nsdi14, par-atc19, ramcloud-sosp11}.
Moreover, to enable fault recovery, they continuously 
persist data to non-volatile storage. 
To decide when an update   in one node should reach the replicas
in other nodes, 
 these systems use different
memory consistency models~\cite{seq-vs-lin-cons-acm94}; to decide when the replicas in the multiple nodes
should be persisted,
these systems use different memory persistency models~\cite{persistency-isca14, minos-hpca24}. 

Most  distributed storage systems, such as ZooKeeper~\cite{zookeeper-atc10},
FaRM \cite{farm-nsdi14}, and RAMCloud~\cite{ramcloud-sosp11} are 
{\em leader-based}. This means that, while client reads can
be processed by any of the   nodes in the cluster, client writes can only be processed
by a single designated node called the {\em Leader}. As the leader processes a write, it 
ensures that the replicas in all the nodes are correctly updated and  
correctly persisted. 

A related approach used in transactional~\cite{dsn01, srds02} and State Machine Replication  (SMR)~\cite{icdcs17} systems
relies on support for atomic broadcast.
Each client broadcasts each transaction or operation
to all the nodes in the cluster, and transactions/operations are totally ordered through a Global
ID. Each node then applies each transaction/operation locally. 
While this approach simplifies recovery, it is costly, as it requires additional software support for the broadcast primitive, usually through consensus~\cite{icdcs17}. 

A recent, higher-performance approach is to organize the cluster as a
{\em leaderless} system~\cite{epaxos-nsdi21, hermes-asplos20, minos-hpca24}. In this approach, both client reads
and writes can be processed by any individual  node in the cluster. A client sends the write request to a single node, 
which 
enforces that all the replicas are updated  and persisted correctly. Leaderless systems are attractive, as they 
can attain superior performance and are 
intrinsically scalable. However, their implementation is   challenging, as the system must support multiple
concurrent writes initiated by different nodes, while updating replicas and persisting
updates following the correct consistency and 
persistency requirements.

Recently, in   distributed systems, ensuring resilience has become an increasingly
growing concern~\cite{apta-dsn23, bluewaters-dsn14, Fault3, exaresilienceUpd-supercomp14,capa-nsdi24}. Both software and hardware failures are becoming more 
frequent. 
To recover from faults, there are a number of proposed  techniques, 
which use
logging~\cite{singlenode-osdi14, adaptivelogging-sigmod16, 
pacman-sigmod17, skyros-sosp21}, replication~\cite{hermes-asplos20, epaxos-nsdi21},   erasure coding~\cite{paritycheck-fast14},
or a combination of them. However, in today's environment, 
it is crucial to deploy light-weight mechanisms that recover a system 
very quickly 
and with only a small impact on the live nodes' throughput.

The most efficient way to recover after the failure of a system component (e.g., 
a node) is to perform {\em online  incremental 
recovery}. Online means that the recovery is performed while the workload
continues executing; incremental means that the recovery consists of re-applying
only the state changes that were missed since the failure. To support online  incremental 
recovery, two mechanisms are required. First, the system needs to keep a 
record of the updates that were completed since the failure and apply them to
the recovering node. Second, these operations need  to be supported while  
the healthy nodes continue processing the reads and writes of the workload.

Unfortunately, no {\em leaderless} system supports this form of recovery.
For example, the state-of-the-art Hermes~\cite{hermes-asplos20} leaderless
system performs online recovery, but recovery requires a  
full database transfer to the recovering node, which can take a long time. 
State-of-the-art leader-based systems such as ZooKeeper~\cite{zookeeper-atc10} perform online incremental recovery.
However, 
their recovery  
works with the simpler, lower-performing
{\em leader-based} protocols.

In addition, recovery in ZooKeeper is not as efficient as it is 
needed by the increasing resilience concerns of cloud clusters.
First, during recovery, the leader node performs a costly access to
its log in persistent storage to retrieve the set of 
updates missed by the recovering node. Second,
this resulting set of updates, which is sent to and persisted in
the recovering node, can contain redundant entries---i.e., multiple
updates to the same record. 
Finally, during recovery, the 
recovering node does not participate in the protocol of the ongoing writes,
which further slows down recovery.

To address these limitations, this paper proposes {\em \name}, 
{\em the first online, incremental recovery scheme for 
leaderless distributed systems}.  
\name{} is a non-transactional system optimized for minimal recovery latency and low throughput impact.
In addition to operating in   leaderless environments, \name{}  introduces 
three techniques for low overhead recovery 
that could also be used by a state-of-the-art
leader-based  recovery scheme  such as ZooKeeper.
First, \name{} provides to the recovering node all the
updates that the latter missed with very low overhead. 
We call this approach {\em Proactive} recovery, in contrast to  {\em Reactive} recovery
in ZooKeeper. Second, the 
provided updates 
contain  no redundant updates---reducing the overheads of transfer and persistency.
Finally, during recovery, the recovering node 
remains active, participating in the protocol of the ongoing writes,
which speeds-up execution.

This paper presents \name{}'s algorithms for the combination of
Linearizable consistency and five
memory persistency models: Synchronous, Strict, ReadEnforced,
Eventual, and Scope.

We implement \name{} on a 6-node Intel cluster with 7 cores per node. 
The results
show that \name{} is very effective. In  
experiments running TAOBench applications~\cite{taobench-vldb22} under Linearizable consistency
and Synchronous persistency, 
\name{} 
recovers the cluster after 
a node failure in
120 milliseconds 
on average, while reducing the throughput of the running workload by 
an  average of only 8.7\%. 
In contrast,
the state-of-the-art
recovery scheme for  leaderless
systems is non-incremental, and needs 360 seconds
to recover,  reducing the throughput of the 
system 
by 16.2\% on average. 
Further, compared to an incremental recovery scheme for a
{\em leader-based} system, \name{} reduces the average recovery 
latency by 20.7$\times$
and the throughput degradation by 62.4\%.

The contributions of this paper are as follows:

\noindent $\bullet$
 \name{}, the first online incremental recovery scheme for 
 leaderless distributed systems.

\noindent $\bullet$
The \name{} design  for 1 consistency and 5 persistency models. 

\noindent $\bullet$
 The \name{} implementation on a 6-node distributed machine. 

\noindent $\bullet$
 An evaluation of \name{}. 
\section{Background}\label{sec:background}

\subsection{Distributed Data Persistency Models}
\label{sec:ddp-models}
Distributed storage systems enhance performance, availability, and resilience by storing  copies of a data record, known as replicas, in multiple nodes. 
The order in which updates are applied to the replicas is governed by the {\em consistency model}.
Additionally, the order in which updates to a replica in
a node become persisted to the durable storage in that node  is defined by the {\em persistency model}. 

The Distributed Data Persistency (DDP) model~\cite{minos-hpca24} of a distributed system is the   combination  of  the consistency and the persistency models of the system.
A DDP model is characterized by two 
concepts:
the {\em Visibility Point}, which is when an update becomes observable, 
and the {\em Durability Point}, which is when an update becomes durable.
The former is dictated by the consistency model, while   the latter by the persistency model. 
The DDP framework allows for various combinations
of $<$consistency, persistency$>$ models.

This paper examines {\em recovery} in the DDP framework.
We consider DDP models that combine the intuitive
Linearizability consistency model~\cite{partnetw-acm85, lin-disc16, lin-acm16, hermes-asplos20}
 with one of five persistency models: Synchronous~\cite{minos-hpca24}, 
 Strict~\cite{Talpey2019}, 
 Read-Enforced~\cite{cad-fast20}, Eventual~\cite{minos-hpca24}, 
 and Scope~\cite{minos-hpca24}. 
 
\subsubsection{Communication in DDP}\label{sec:ddp-model-defs}

The node that initiates a write request is named the \textit{Coordinator}, and the rest of the  nodes are \textit{Followers}. In a write operation, the Coordinator sends an invalidation (INV) message to the Followers, passing the new data.
The Followers acknowledge the INV with an ACK message, confirming that 
they have applied the update 
based on the supported consistency, persistency, or both. Subsequently, when the Coordinator receives all the  ACKs, and after it has applied the update locally,
it  
sends a validation (VAL) message to the Followers, signaling the transaction's completion. 
Depending on the DDP model,
ACK or VAL messages
may have a consistency and a persistency flavor (e.g., ACK\_C and ACK\_P).

\subsubsection{Model Definitions}
\label{defs}
We consider these models:

\noindent
\textbf{Linearizable Consistency (Lin)} 
ensures a total order  of read and write operations to the volatile state across all nodes. 
In a write, a response  is sent back to the client only after updating the volatile state of all the replicas. 

\noindent
\textbf{Synchronous Persistency (Synch)} requires the write to be made durable in a node as soon as the local volatile replica is updated. Combining $<$Lin, Synch$>$ means that a  response is issued to the client immediately after all the replicas' volatile and non-volatile state is updated. The Coordinator issues the response upon receiving all the ACKs and persisting locally.

\noindent
\textbf{Read-Enforced Persistency (REnf)} 
necessitates that a write be persisted in all the replicas before it is read. In the $<$Lin, REnf$>$ combination, the client receives a write response once all the volatile replicas are updated (indicated by the Coordinator
receiving all ACK\_Cs). When the Coordinator receives all  ACK\_Cs and ACK\_Ps, it knows that all
the replicas are updated and persisted. Then, it sends a VAL message to all the Followers,
which then allow reads to the record.

\noindent
\textbf{Eventual Persistency (Event)} 
 stipulates that writes to replicas should be persisted
 {\em eventually}, without delaying 
 subsequent reads or writes to the same data.
 With $<$Lin, Event$>$, the client receives the write response after all the volatile replicas are updated. The replicas will be persisted in the future.

\noindent
\textbf{Scope Persistency (Scope)} 
 uses the concept of \enquote{scopes}, which are sets of read and write operations. Messages are tagged with an additional \enquote{sc} to indicate their scope, like [INV]sc.
 A client indicates the end of a scope with a  [PERSIST]sc command. Under this model, a response to a [PERSIST]sc is sent back to the client only after all the writes within the scope are persisted. 
 When $<$Lin, Scope$>$ is used: a) the response to each write within a scope is returned when the write updates all volatile replicas, and b) the response to [PERSIST]sc is returned only when all the writes in the scope have been applied to both the volatile and non-volatile state of all replicas. 
When the Coordinator receives the ACK\_P for the
[PERSIST]sc from all the Followers, it
knows that all the writes in scope \enquote{sc} 
are persisted. It then informs the client, and
sends a  [VAL\_P]sc  to all Followers.

\noindent
\textbf{Strict Persistency (Strict)} 
is the most stringent model. It dictates that a write must be persisted in all the replica nodes before the client receives the write response. This can occur even before the volatile state of the replicas is  updated. 
It separates consistency and persistency with two types of ACKs (ACK\_C and ACK\_P) and VALs (VAL\_C and VAL\_P). 

\subsection{Baseline Leaderless Protocol} 
\label{sec:background:timestamps}
\label{sec:baseline-leaderless}

The baseline leaderless protocol implements a write operation with the  message sequence described in
Section~\ref{sec:ddp-model-defs}, which was introduced by Hermes\cite{hermes-asplos20}.
Since any node can initiate a write, to ensure correct ordering, 
the protocol, like Hermes,
relies on Lamport's Logical Timestamps~\cite{clocks-acm78}.
Specifically, each record in the datastore has a timestamp, which is
a tuple $<$node\_id, version$>$, where 
{\em  version} increases monotonically at every update,
and  {\em node\_id} is the 
Coordinator that generated that version.
Each write request carries a timestamp, where the 
{\em  version} is set by fetching the current   version of the record and incrementing it by one.
The order of writes to a given record is determined by their timestamps. 
When comparing two writes,
the one with the higher timestamp version 
 is considered newer. If two version numbers are equal, then the write with the higher node\_id is considered newer. 
 
 Only newer writes are applied to a node's memory.
 Updates with lower timestamps are
discarded upon arrival.
Overall, all
healthy nodes that participate in a set of committed updates will end up
with
the same final versions of all records, even if intermediate operations are
reordered or retransmitted. 

\subsection{ZooKeeper's Online Incremental Recovery}
\label{zookeeper_recov}

ZooKeeper uses a leader-based protocol~\cite{zookeeper-atc10}. 
When the leader performs a write, as part of the transaction,  
a quorum of nodes in the system  record  the write in a local 
log in persistent storage. For recovery, ZooKeeper
uses a combination
of checkpointing and   incremental updates. When a node fails, the leader uses
its own log to bring up a replacement node.

While the recovery is online and incremental, it has three limitations.
First, the recovery starts-off with the replacement (i.e., {\em Recovering})
node sending to the leader the ID
of the last record ({\em zxID}) that the recovering node finds
in its log. When the leader receives the zxID, it accesses its own 
log to find zxID and retrieve  all the logged updates since that point. As
this operation takes time, we call it a {\em Reactive} recovery.

Second, the set of updates that the leader extracts from its log  may  contain
multiple updates to the same record. All the updates to the record except
the last one are effectively redundant. 
Still, the leader sends all the 
updates to the recovering node, which applies them to its own persistent  log.

Third, during recovery, the recovering node applies the updates to its log, but 
does not participate in  ongoing transactions. As the leader processes 
new 
writes and other nodes log them, the leader buffers the updates intended for the 
recovering node. Then, it sends them to the recovering node when the latter has finished logging the previous 
batch of updates. This process is suboptimal and may repeat multiple times.

\subsection{Why Not Adapt Leader-based Recovery for Leaderless?}
\label{adapt_recov}

There are multiple reasons why leader-based fault detection/recovery cannot be adapted for leaderless systems.
First, leader-based architectures use the leader node as a serialization point, 
to streamline write processing and recovery. In leaderless systems, multiple nodes may concurrently 
process different writes for the same record (even during recovery), needing careful versioning for consistency. While one could use a broadcast primitive to enforce global ordering, it adds
overhead.
Second, 
in leader-based systems, failed non-leader nodes can be easily discarded.
In contrast, leaderless systems require   more careful failure-handling, as each node can serve  as   Coordinator and as Follower. Finally, leader-based systems need a  process of
leader election when the leader node fails~\cite{rtzk-leaderelec-ecs21}. This is unnecessary in leaderless systems.
\section{Motivation}\label{sec:motivation}

To motivate the need for  fast recovery after a  node failure, 
we consider the recovery of a state-of-the-art leaderless
system like Hermes~\cite{hermes-asplos20},
which involves copying the complete key-value store from a healthy node
to a new node that replaces the one that failed. 
We implement the leaderless Hermes protocol in
a 2-node cluster  
with Xeon E5-2450 CPUs (Section~\ref{sec:methodology}).
We refer to the  new  node
as the  {\em Recovering}  node, and the healthy one that provides the data as the
 {\em Recoverer} node.
We use one-sided 
RDMA writes to transfer  data from   recoverer   to   recovering
node. 

We consider  three 
environments. In 
\textit{DRAM-only}, the data is moved from the 
recoverer's DRAM memory to the recovering's DRAM memory.
This is the design assumed in Hermes. However, a likely more realistic environment would use
persistent memory (PM) like  
Intel Optane~\cite{characterizing-pmem-micro20}\footnote{Although Optane is discontinued, other alternatives can be used~\cite{postoptane-dimes23, SNIA2024_EmergingMemories}.
For example, Micron demonstrated a 32\,Gb (4\,GB) NVDRAM device at IEDM~\cite{nvdram-micron-iedm23}, Everspin ships standalone STT-MRAM parts, Panasonic and STMicroelectronics continue to advance ReRAM and embedded PCM, and CXL explicitly supports emerging persistent media~\cite{SNIA2024_EmergingMemories}.}  
for reliability. 
In this case, we consider two designs. 
In {\em DRAM+PM}, the data is moved from the 
recoverer's PM to its DRAM, then sent via RDMA to the
recovering's DRAM, and then persisted to the recovering's
PM. In contrast, in  {\em PM-only}, the data is sent via
RDMA directly from the recoverer's PM to the recovering's
PM. We believe {\em DRAM+PM} is preferable over {\em PM-only}
because it avoids
any inconsistency between DRAM and PM data.

\begin{figure}[t]
    \centering
    \includegraphics[width=0.4\textwidth]{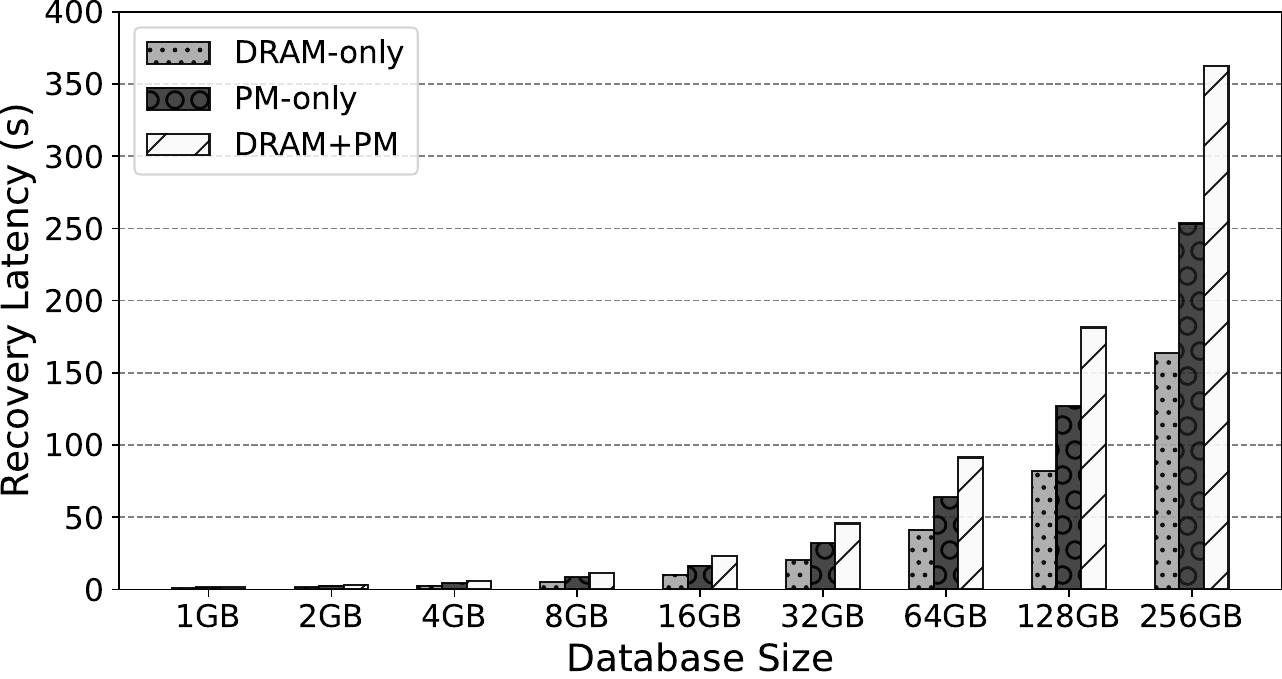}
      \vspace{-2mm}
    \caption{Recovery latency in a state-of-the-art leaderless system.}
    \label{fig:hermes-recov}
    \vspace{-5mm}
\end{figure}

Figure~\ref{fig:hermes-recov} shows the latency of the data transfer, which is the
recovery time, as the 
size of the key-value  store changes between 1GB--256GB. 
As the database size grows, the
recovery latency increases. For a 256GB database,   
\textit{PM-only}  takes 254s, while
  \textit{DRAM+PM}  takes 360s. These are significant overheads.

\section{Failure Model}
\label{sec_fault}
\label{sec:overview:fail-model}

Cloud systems are expected to provide high availability.
Even brief outages are highly undesirable~\cite{SpannerTrueTimeCAP, outagesGoogle-2006}, 
as they can lead  to potential financial losses of millions of dollars\cite{productionBugs-hotOS19}.
In this paper, we assume a scenario where a node in the
cluster fails or disconnects, but its durable state remains intact 
and accessible after a short downtime (e.g., seconds to  minutes). 
On regaining access, the durable state can be  either in the original  failed node, now back online, or in a replacement node. 

The first
scenario (i.e., durable state in the original node) can be the result of various
hardware or software  faults~\cite{cloudFaults-esec22, cloudFaults-issre22}.
Example hardware faults are internal CPU errors, CPU frequency drops, I/O latency anomalies,
overheating, or service exceptions. Example software faults are kernel crashes, firmware issues, application bugs, misconfigurations, memory/resource
leaks, or software incompatibilities. 
This scenario can also be the result of regular fleet operations like rack maintenance 
or software/BIOS updates.
These issues are frequently addressed through reconnection or 
reboot~\cite{cloudFaults-esec22, cloudFaults-issre22, hpc-failures-tdsc10}.

The second scenario (i.e., durable state found in a replacement node) appears 
when hardware replacement is needed to
mitigate infrastructure-related errors~\cite{LLM-datacenter-failures-nsdi24}, and may require  the involvement of an operator~\cite{HwFails-OperatorInvolvement-storage18, HwFails-OperatorInvolvement-sosp23}. Storage systems are generally resilient to failures, as they include 
technologies like battery-backed DRAMs~\cite{batteryDRAM-isca17, bbb-hpca24} and eADR~\cite{eADR-pact22}. On a node failure, they
can be hot-plugged~\cite{premio-2022}, allowing their transfer to a healthy node added to the cluster. 
Finally, datacenters often utilize dedicated storage servers~\cite{storageServers-hotcarbon24}, which ensure  that remote storage remains intact and accessible by a replacement node in the event of a compute node failure.

Many of these errors are resolved within minutes. For example,
the median restart time due to infrastructure issues is 8.6 minutes~\cite{LLM-datacenter-failures-nsdi24},
while the median downtime following a crash  is  
11.43 minutes, with about half of these restarts classified as 
unplanned~\cite{downtime-srds99,datacenters-barroso-book-2018}. 
Further, thanks to
link redundancy,
link failures are typically autonomously resolved without the need for an
operator 
within a 5-minute time frame~\cite{linkFailures-sigcomm11}. 

In these scenarios, (i) the ability of the software to access 
the durable state again as it was before the 
failure, and (ii) the short duration of the downtime,  
provide  an opportunity to use high-performance online  incremental recovery. 
The healthy nodes continue executing the application as usual---both while the
failed node is down and while a node is recovering. 

In cases where the durable state of a node is lost or remains inaccessible for an extended period, incremental recovery is not feasible. Then, the persistent state of the recovering node must be updated with 
the full copy of the datastore.

We   assume that one node in the system runs a  Configuration Manager (CM)  process that regularly
sends  heartbeat messages to all the  other nodes~\cite{zookeeper-atc10, heartbeat-osdi20}. If a node fails to respond within a timeout period, a failure is flagged. 
In line with prior work \cite{zookeeper-atc10, hermes-asplos20, craq-atc09, chain-osdi04}, we assume the CM is fault tolerant. The CM periodically notifies a dedicated distributed service. If the service does not hear from the CM node for a while, it relaunches a new instance of the CM node.
This is a well-established  approach to detect node failures.

\label{zk-onlyforCM-update}
Other types of failures beyond a node crashing or getting disconnected,
such as byzantine failures~\cite{byzantine-springer03} or permanent
network partitions, 
are not supported.  
We also assume that the system 
addresses
silent   data corruption by using 
existing
RAS features and error-correcting codes in DRAM and network. 

\label{mentionConcFailsEarlier-update}
Our work primarily assumes a single failure in the system at any given time,
which aligns with prior work~\cite{viewstamped-mit12}. However, we extend our discussion to concurrent failures in Section~\ref{sec:concurrent-fails}. 

\section{\name{}: Recovery in Leaderless Systems}\label{sec:algorithms}

\name{} is a general and high-performance online incremental 
recovery scheme for leaderless distributed systems. 
It
 builds on top of an
implementation of MINOS~\cite{minos-hpca24}  
and supports multiple 
memory persistency models. 
This section describes \name.

\subsection{Timeline of Fault Detection and Recovery}\label{sec:detect_reconfig}

\begin{figure}[t]
    \centering
    \includegraphics[width=0.45\textwidth]{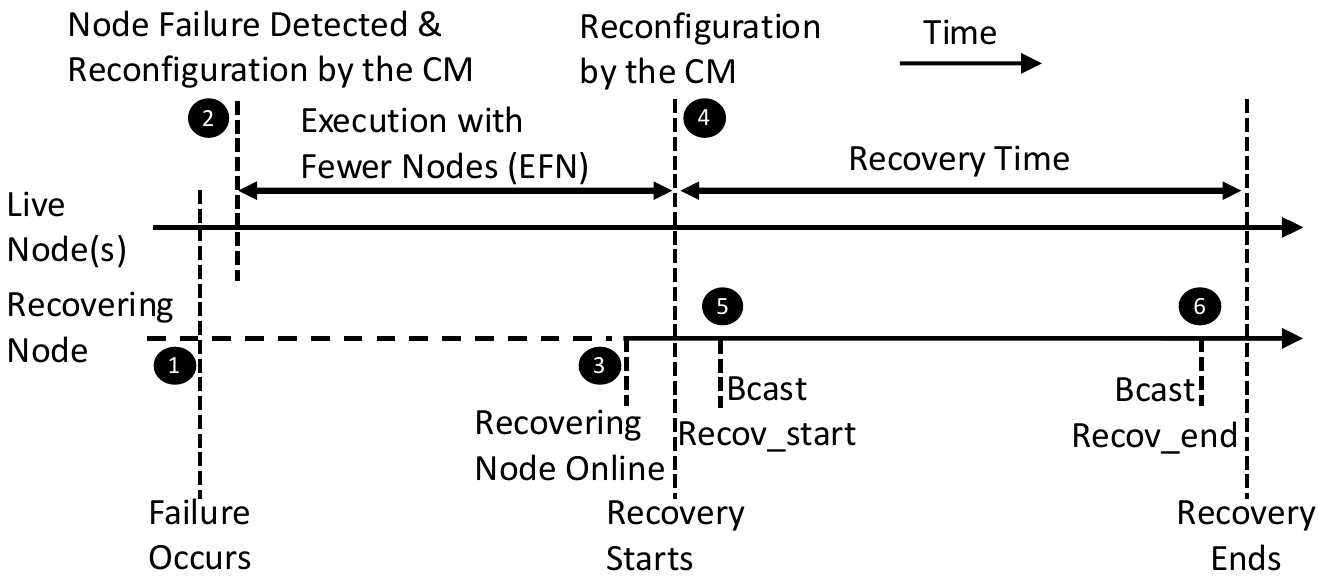}
    \caption{Fault detection, reconfiguration, and recovery.}
    \label{fig:recov-timeline}
   \vspace{-3mm}
\end{figure}

To put the operation of \name{} in context, Figure~\ref{fig:recov-timeline} shows
the timeline of fault detection and recovery. Before the \name{} recovery algorithm executes, two steps  take place: fault detection and system reconfiguration to isolate
the faulty node. How to perform these steps is not a contribution of this paper, as they have
been detailed elsewhere~\cite{farm-nsdi14, farm-sosp15, hermes-asplos20}. 
Essentially, faults are detected by the CM through timeouts, and
reconfiguration is performed by the CM  by removing the failed node
from the list of live nodes, and broadcasting the new list to all the live nodes.

After reconfiguration~\circleint{2}, the live nodes continue executing the workload, now without communicating
with the failed node. This period is called the
{\em Execution with Fewer Nodes} (EFN).
Then, at some point, a recovering node $R$ is identified by the CM~\circleint{3}. 
As indicated in Section~\ref{sec_fault}, $R$ may be the node that failed
or a new one 
but, in either case, it contains the durable state of the failed node at the point
of failure.
After $R$ is 
ready to recover, the
CM  changes the configuration again to include $R$ 
in the list of live nodes, and 
broadcasts it to all the live nodes~\circleint{4}. The {\em Recovery time} now starts.

\subsection{Key Ideas and Structures for the \name{} Recovery}
\label{sub_def}

To perform its recovery, \name{} relies on extended definitions of the 
Durability and Visibility points, and on two software structures:
the \textit{\LDP{}} list and the
\textit{\newD} buffer. We describe them next.

\subsubsection{Durability Point}
\label{sec:alg:durability-point}

Prior work~\cite{minos-hpca24}
defines the Durability Point (DP) of an 
update in a distributed system as the point when 
the update is pushed to non-volatile storage.  
In this paper, we extend the concept by defining a \textit{Local Durability Point} (LDP)  for 
each node. This is the time when the update is pushed to local non-volatile
storage in that  node---e.g., a non-volatile memory (NVM) log. 
Also, we define the \textit{Global Durability Point} (GDP) as the time when the update is persisted in all replica nodes. The GDP is 
observed by different nodes at different times. 
Since
an update has a Coordinator  and multiple Follower nodes, 
we define a \textit{Coordinator GDP} (CGDP) and per-follower 
\textit{Follower GDP} (FGDP). 

\begin{figure}[t]
    \centering
    \includegraphics[width=0.38\textwidth]
    {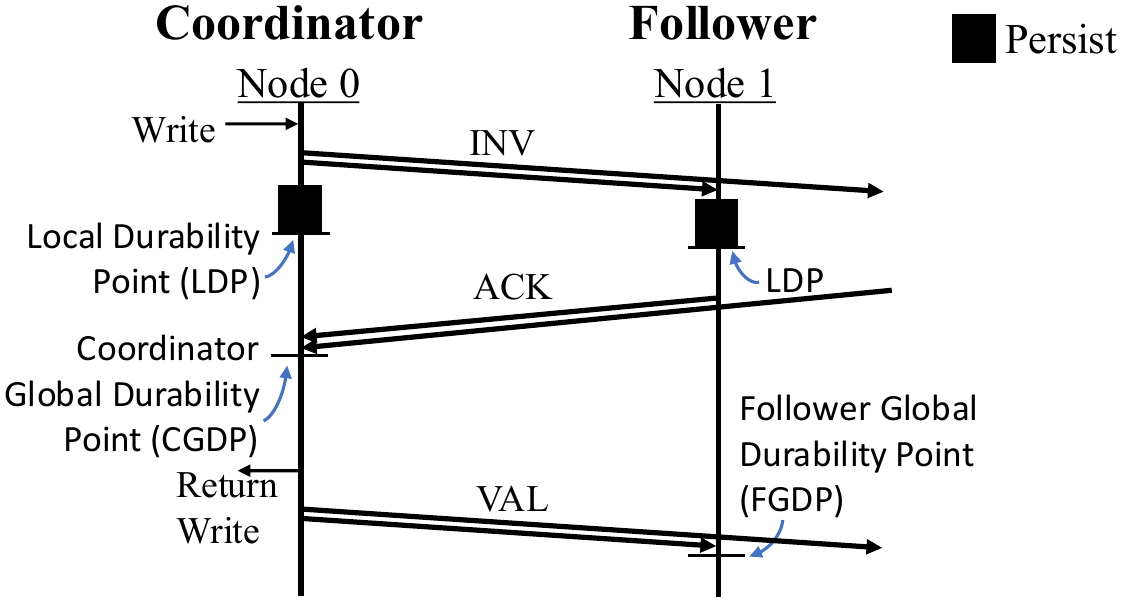}
    \vspace{-2mm}
    \caption{Durability points in an update under $<$Lin, Synch$>$.} 
    \label{fig:dur-point}
    \vspace{-6mm}
    
\end{figure}

Figure~\ref{fig:dur-point} shows the transaction for an update as described
in Section~\ref{sec:ddp-model-defs}. It corresponds to the $<$Lin, Synch$>$ model.
We show a single Follower but there are multiple Followers since, in this paper, we will 
assume that a record is replicated in all Followers. 
The figure shows the LDPs, the CGDP, and the FGDPs.
The Coordinator reaches its LDP after it 
 has persisted the update locally.
 Further, it reaches the CGDP after it has persisted the update locally and has received
 the ACKs from all nodes. A Follower reaches its LDP after it 
 has persisted the update locally. It reaches the 
 FGDP after it has received the VAL  from the Coordinator.
Section~\ref{sec:alg:iaso-pers-models} describes the LDP, CGDP, and FGDP of the different
models considered.

\subsubsection{Visibility Point}\label{sec:alg:visibility-point}

Prior work~\cite{minos-hpca24}  
defines the Visibility Point (VP) of an 
update as the point when the update becomes
available for consumption. It is equivalent to the DP
except that VP deals with updates to the {\em volatile} storage. Hence, we also
define a Local Visibility Point (LVP) per node, a Coordinator Global Visibility Point (CGVP),
and a per-follower Follower Global Visibility Point (FGVP), in a similar manner. 
Section~\ref{sec:alg:iaso-pers-models} describes the LVP, CGVP, and FGVP of the different
models considered.

\subsubsection{\textit{\LDP{}} List}
\label{sec:alg:lpded-tx}

In a Coordinator or Follower node, a  {\em Partially-persisted} update  is one
 that has been persisted in the local NVM log
 (and, therefore, has reached the LDP), 
but has not yet reached the CGDP or FGDP, respectively. 
To identify these updates, we use the {\em GDP Flag}.
Each entry in the local NVM log has a {\em GDP Flag}, which is set when
the update finally reaches the FGDP (or CGDP).

If a node fails while an update has a clear
GDP Flag, the update presents a dilemma during recovery, as the 
recovering node does not know whether the  update  reached the GDP in all the
other nodes. Hence, during \name's recovery,
the recovering node compiles 
a list of
all the partially-persisted updates in its NVM log.
The  list is called the
\textit{\LDP{}} list.

\subsubsection{\textit{\newD} Buffer}\label{sec:alg:newdatabuf}

While a node is down, all the other nodes continue executing the application. Each of them
keeps buffering all the application updates in a volatile buffer in local memory called 
\textit{\newD}. This buffer will be used during recovery. We allow each live node 
to generate its own buffer for redundancy and  recovery simplicity.

The set of updates in \newD{} is not the same as in the 
NVM log. First, the NVM log is periodically recycled by pushing
its entries to
longer-term permanent storage. Second, 
only updates that had their GDP Flag set go to \newD. Finally,
in  \newD, newer updates to a record 
overwrite earlier updates to the same record. 

\label{sec::alg::newDThreshold}
To prevent excessive DRAM usage, 
we limit   the size of the \newD{} buffer. 
Once this limit is exceeded, the buffers are
discarded and recovery falls back to non-incremental recovery. 

\subsection{\name{} Recovery Algorithm}\label{sec:alg:recovery-algorithm}

After the CM changes
the configuration for the second time, adding the recovering node $R$,
and all the nodes are informed (\circleint{4} in Figure~\ref{fig:recov-timeline}), recovery starts.
The idea behind the recovery is simple.  $R$ needs to ask a statically determined node
(called its \emph{Buddy} node) 
whether the stores in $R$'s 
\textit{\LDP{}} list where finally
globally persisted. Further, $R$ needs to get from its Buddy node the list of updates 
in the Buddy node's \textit{\newD} buffer and apply them locally. During all this process, $R$ needs to
participate in the commit of application updates initiated by other nodes---although $R$
does not  accept direct client requests until the recovery is over. Finally, 
$R$ needs to synchronize with all the other live nodes 
to initiate \circleint{5} and
to terminate \circleint{6} the recovery.

\begin{figure}[t]
    \centering
    \includegraphics[width=0.4\textwidth]{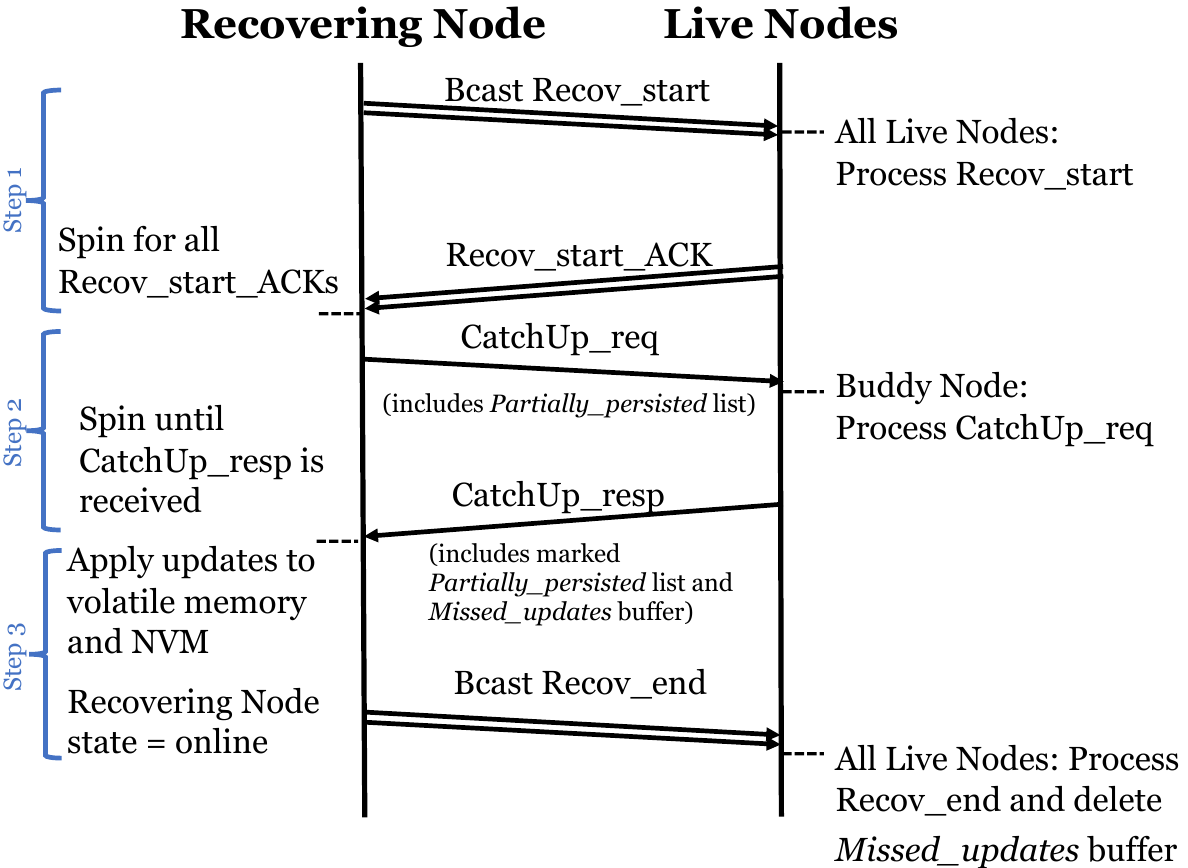}
    \vspace{-2mm}
    \caption{\name{} recovery algorithm.}
    \label{fig:recov-recovering-node}
    \vspace*{-.4\baselineskip}
    \vspace{-1mm}
\end{figure}

Figure~\ref{fig:recov-recovering-node} depicts the recovery algorithm for the  $<$Lin, Synch$>$ model. Later, in Section~\ref{sec:alg:iaso-pers-models}, we consider the other models. Table~\ref{tab:recovery-msgs}
describes the messages exchanged in the algorithm. The algorithm has 
three steps.

\begin{table*}[t]
    \centering 
     \begin{scriptsize}
     \caption{Times when an update reaches the different visibility and durability points for different
    models. In the table, \enquote{*} does not correspond to the update but, instead, to the
PERSIST command that finishes the scope of the corresponding update.}  
    \vspace*{-.6\baselineskip}
      \vspace{-1mm}
\begin{tabular}{|l|l|l|l|l|l|l|}
        \hline
        \rowcolor[HTML]{EFEFEF}
\textbf{Model} & \textbf{Reach LVP} & \textbf{Reach CGVP} & \textbf{Reach FGVP} & \textbf{Reach LDP} & \textbf{Reach CGDP} & \textbf{Reach FGDP} \\ \hline
$<$Lin, Synch$>$ &  & All ACKs received & VAL received &  & All ACKs received & VAL received \\  \cline{1-1} \cline{3-4} \cline{6-7}
$<$Lin, Strict$>$ &  &  & VAL\_C received &  &  & VAL\_P received \\ \cline{1-1} \cline{4-4} \cline{7-7}
$<$Lin, REnf$>$ &  &  & VAL received & \multirow{-3}{*}{\begin{tabular}[c]{@{}l@{}}Updated \\ local \\ NVM\end{tabular}} & \multirow{-2}{*}{\begin{tabular}[c]{@{}l@{}}All ACK\_Ps \\ received\end{tabular}} & VAL received \\ \cline{1-1} \cline{4-7}
$<$Lin, Event$>$ &  & \multirow{-3}{*}{\begin{tabular}[c]{@{}l@{}}All ACK\_Cs \\ received\end{tabular}} & VAL\_C received & --- & --- & ---\\ \cline{1-1} \cline{3-7}
$<$Lin, Scope$>$ & \multirow{-5}{*}{\begin{tabular}[c]{@{}l@{}}Updated\\ local\\ volatile\\ memory\end{tabular}} & \begin{tabular}[c]{@{}l@{}}All {[}ACK\_C{]}sc's \\ received\end{tabular} & {[}VAL\_C{]}sc received & \begin{tabular}[c]{@{}l@{}}Updated \\ local NVM*\end{tabular} & \begin{tabular}[c]{@{}l@{}}All {[}ACK\_P{]}sc's\\ received*\end{tabular} & \begin{tabular}[c]{@{}l@{}}{[}VAL\_P{]}sc \\ received*\end{tabular} \\ \hline
\end{tabular}
    \label{tab:iaso-terms}
    \end{scriptsize}
    \vspace{-3mm}
\end{table*}

\noindent
\textbf{Step 1:} 
As soon as the CM completes the second reconfiguration,   $R$ {\em participates} as a Follower
in the protocol of all the  updates initiated by other nodes. However, 
it cannot yet take any client requests.
We say that $R$ acts as a 
\textit{ShadowReplica}~\cite{hermes-asplos20}.

In this first step, $R$
broadcasts a {\em Recov\_start} message to all live nodes. Then, it
uses its local NVM log to compile its \LDP{} list. 
In practice, in most cases, the \LDP{} list is empty.
Finally, all  live nodes respond with
a {\em Recov\_start\_ACK} message. 

\noindent
\textbf{Step 2:} $R$ sends a {\em CatchUp\_req} 
message that includes its \LDP{} list to its Buddy node. The Buddy node checks 
the timestamp $T_{in}$ of each entry in such \LDP{} list against the timestamp 
$T_{curr}$ of the corresponding record
in the Buddy node's volatile datastore state.
Three cases are possible. If $T_{curr}$ is equal to $T_{in}$, the 
entry in the \LDP{} list is marked
as globally committed. If $T_{curr}$ is less than $T_{in}$, the entry  in the \LDP{} list 
is not marked 
because the update never  committed. Finally, if 
$T_{curr}$ is higher than $T_{in}$, a later update to the record has committed.
Since such update must be in the \newD{} buffer, there is no need to mark
the entry  in the \LDP{} list as committed.

\begin{table}[t]
    \centering
    \caption{Messages exchanged in the recovery algorithm.}
    \vspace{-2mm}
    \begin{scriptsize}
        \begin{tabular}{|l|l|}
            \hline
            \rowcolor[HTML]{EFEFEF}
            \textbf{Message} & \textbf{Description} \\ \hline
            \multirow{2}{*}{Recov\_start}
            & Broadcasted by the recovering node $R$ to all live   nodes  to \\
            & initiate the recovery.\\ \hline 

            \multirow{1}{*}{Recov\_start\_ACK}
            & Sent by live nodes to $R$ as a  response to Recov\_start.\\ 
            \hline
            \multirow{2}{*}{CatchUp\_req}
            & Sent by $R$ to the Buddy node   to request the updates  \\
            & it has missed. It includes   its \LDP{} list.
            \\ \hline
            \multirow{3}{*}{CatchUp\_resp}
            & Sent by the Buddy node to $R$. It includes $R$'s \\
            &  \LDP{} list marked by the Buddy node and\\
            &  the Buddy node's \newD{} buffer.\\ \hline
            \multirow{1}{*}{Recov\_end}
            & Broadcasted by $R$ to all live  nodes to complete the recovery. \\ \hline
        \end{tabular}
    \end{scriptsize}
    \label{tab:recovery-msgs}
    \vspace*{-1\baselineskip}
    \vspace{-2mm}
\end{table}

Then, the Buddy node responds to $R$ with a {\em CatchUp\_Resp} message with: 1)
the marked \LDP{} list and 2) the contents of the Buddy node's \newD{} buffer. 

\noindent
\textbf{Step 3:} 
When $R$ receives {\em CatchUp\_Resp}, it applies locally 
the 
marked updates from the received
\LDP{} list, and the updates from the received \newD{} buffer. 
The updates are applied to both volatile memory and NVM to bring both of them up-to-date.

Then,
$R$ changes its state to fully online, which allows it to take
 client requests. It broadcasts a {\em Recov\_end}  
message to all live nodes to mark the end of the recovery. When a live node
receives the message, it deletes its \newD{} buffer.

\subsection{Recovery Issues}\label{sec:alg:iaso-recov-timeline}
\vspace{-1mm}

In \name{}, no  committed  update is lost. At the time a node fails (\circleint{1} in Figure~\ref{fig:recov-timeline}), 
its local NVM log has an accurate record of 
what updates have been globally
persisted (i.e., the entries with GDP Flag set) or at least locally  persisted (i.e.,
those with GDP Flag 
clear). 
In addition, any update initiated from a  live node after the node failure   but
before the first CM-triggered reconfiguration
{\em cannot complete}. The reason is that it requires ACKs from all Followers, including 
the failed node. The communication with the failed node times out, and the write cannot 
complete. Eventually, the CM discovers the failure. After reconfiguration, during the EFN period,
all updates are saved in the live nodes' \newD{} buffers.

During recovery, $R$ proceeds to apply the updates received in the {\em CatchUp\_Resp} 
message from the Buddy node. However, it rejects those that have already
become obsolete due to a later update to
the same record during its execution as a \textit{ShadowReplica}.

The longer the EFN period is, 
the larger the 
\newD{} buffer in the Buddy node will grow, and the longer it will take for $R$
to process the
buffer during the recovery. 
The duration of the EFN period depends on
how quickly an $R$ can be brought online and is orthogonal to 
\name{}.

The time between a node  failure and   failure detection is likely to be  small. This
is because it is determined by the timeout of 
a CM heartbeat to the  failed node.

\subsection{Impact of Consistency-Persistency Models}
\label{sec:alg:iaso-pers-models}
\vspace{-1mm}

Table~\ref{tab:iaso-terms} shows when 
the Local, Coordinator Global, and Follower Global Visibility Points (LVP, CGVP, and FGVP)
and Durability
Points (LDP, CGDP, and FGDP) 
are reached for an update in the  five 
$<$consistency, persistency$>$
models considered. The terminology 
used follows MINOS~\cite{minos-hpca24}.
For example, in $<$Lin, REnf$>$: 
(i) a node reaches LVP when it updates its local volatile memory;
(ii) a Coordinator reaches CGVP when it receives all the consistency ACKs (ACK\_Cs);
(iii) a Follower reaches FGVP when it receives the VAL;
(iv) a node reaches LDP when it updates its local persistent memory;
(v) a Coordinator reaches CGDP when it receives all the persistency ACKs (ACK\_Ps); and
(vi) a Follower reaches FGDP when it receives the VAL.

Note that $<$Lin, Event$>$ has no defined durability points for an update, 
as it  persists ``eventually''.
Moreover, as shown with an asterisk,  the durability point conditions for $<$Lin, Scope$>$ refer not
to messages  in the update itself but, instead, to messages 
in the  PERSIST  that finishes the scope of the update.
These visibility/durability point considerations are in effect
during normal execution and during ShadowReplica execution. 
Hence, they are in effect 
when updating the \newD{} buffer 
and when persisting state in the local NVM log.

\begin{table*}[ht!]
    \vspace{-2mm}
    \centering
    \caption{Node failures and recovery from them.}
    \vspace*{-.6\baselineskip}
    \begin{footnotesize}
    \begin{tabular}{|l|l|l|l|}
    \hline
    \rowcolor[HTML]{EFEFEF} 
    \textbf{Type of Update Failure} &
    \textbf{\begin{tabular}[c]{@{}l@{}}How the Failure Makes the\\ Update Operation Incomplete\end{tabular}} &
    \textbf{\begin{tabular}[c]{@{}l@{}}Completing the Update during Execution \\with Fewer Nodes\end{tabular}} &
    \textbf{\begin{tabular}[c]{@{}l@{}}What the Recovering Node\\ Does during Recovery\end{tabular}} \\ 
    \hline

    \begin{tabular}[c]{@{}l@{}}Coordinator fails  after \\ sending at  least 1 INV  \\ and before sending all  VALs\end{tabular} &
    \begin{tabular}[c]{@{}l@{}}At least one\\ Follower received\\ INV but no VAL\end{tabular} &
    \begin{tabular}[c]{@{}l@{}}CM reconfigures to remove Coordinator and inform \\ all other nodes.  One or more live nodes that had \\received   INV but no VAL take over  as Coordinator\\ and replay the update\end{tabular} &
    \multirow{2}{*}{\begin{tabular}[c]{@{}l@{}}Apply the update from the \\\newD{} buffer of the \\Buddy Node (if update has \\not become
    obsolete already)\end{tabular}} \\
    \cline{1-3}

    \begin{tabular}[c]{@{}l@{}}A Follower fails\\ before sending the\\ ACK to the Coordinator\end{tabular} &
    \begin{tabular}[c]{@{}l@{}}The Coordinator does not\\ collect all ACKs\end{tabular} &
    \begin{tabular}[c]{@{}l@{}}CM reconfigures to remove Follower\\ and informs all other nodes.\\ The Coordinator replays the update\end{tabular} 
    & \\

    \hline
    \end{tabular}
    \end{footnotesize}
    \vspace{-4mm}
    \label{tab:recov-coord}
\end{table*}

\subsection{Failure Scenarios}\label{sec:fail-scenarios}
\vspace{-1mm}
\name{} considers two  types of failures. The first one includes
message loss, long message delays due to high contention, and intermittent network link disconnections;
the second one includes a node failing or getting  permanently disconnected from
the network. As indicated in Section~\ref{sec_fault}, other failures like 
permanent network partitioning are not considered.

To understand how to handle the supported failures,
recall that, when a 
node receives a client read, it satisfies it locally, while when it receives a client write,
it involves all the other nodes.
If a failure occurs during a read, the read does not complete and the client retries after a 
timeout using a different replica node. This is trivial and so we do not explore it.
Hence, we focus on  failures during a write.
For simplicity, we use   $<$Lin, Synch$>$, but similar ideas apply to the other models.
We consider the two types of failures.

\label{sec:alg:msg-detect}

\subsubsection{Recovering from Message Loss/Delay and Intermittent Network Disconnections}  
\label{sec:alg:msg-loss-wr}

These failures may cause two types of timeouts (Figure~\ref{fig:dur-point}).  The first one 
({\em INV-ACK}) goes off in a Coordinator if the time from when all the INVs have been sent until when
all the corresponding ACKs have been received is over a threshold. The second one 
({\em ACK-VAL}) goes off in
a Follower if the time from when it sends an ACK until when it receives the corresponding VAL is over a threshold. 

When a Coordinator's {\em INV-ACK} timeout goes off, the Coordinator
rebroadcasts the INVs to all Followers,
similarly to Hermes~\cite{hermes-asplos20}. The process is 
potentially repeated multiple times.
We opt for rebroadcasting over sending messages to only
the non-responding nodes
to reset the timers in all the Followers. Each message includes the initial request 
timestamp, which remains unchanged upon replay, facilitating the  handling  of duplicates.

When a Follower's {\em ACK-VAL} times out, the Follower resends
the ACK as in the previous case. 
Overall, 
in both cases, the response will 
eventually be received.

\subsubsection{Recovering from Node Failures and Permanent Network Disconnections}
\label{sec:crash}

These failures are identified through non-responsive CM heartbeats. 
To understand the \name{} recovery, we consider the two possible cases: Coordinator
failure (or disconnect) and Follower failure (or disconnect). Table~\ref{tab:recov-coord} shows, for each of the
two failures, 
(1) how the failure makes the update operation incomplete, (2) how the incomplete operation
completes during the execution with fewer nodes,  
and (3) what  the 
recovering node $R$ does during the recovery. 

A {\em Coordinator failure} occurs when the Coordinator of an update fails after sending 
at least one INV and before sending all VALs---we ignore the trivial case when a 
Coordinator fails before sending any of the 
INVs because, in this case, the client times out and reissues the request to another node.
The operation is incomplete because there is at least one {\em incomplete}
Follower, i.e., a Follower that received an INV and no VAL.
Once the CM has detected the failure, it reconfigures the system to remove the 
Coordinator node and informs all the remaining nodes. At this point, an incomplete Follower
takes on the role of the Coordinator for this update, and replays the update, carrying the original Coordinator's   timestamp. 
\name{} sets the reaction times of the Followers so
that one reacts first. However, it is possible that multiple incomplete Followers react simultaneously, 
which  leads to multiple Followers 
taking on the role of Coordinator for the same update.  
This is correct because a node processes requests 
according to timestamps. Multiple messages with the same timestamp 
induce the same effect as a 
single message, as duplicates are obsolete.  

A {\em Follower failure} occurs when a Follower fails before sending the ACK to
the Coordinator of an update. The
operation is incomplete because the Coordinator does not collect all the ACKs.
Once the CM has detected the failure, it reconfigures the system to remove the 
Follower node and informs all the other nodes. Then, the Coordinator replays the update 
without including the failed
node.

\subsection{Comparison to ZooKeeper's Recovery}
\label{zookeeper_compare}

The \name{} recovery is  more efficient than the
ZooKeeper one described in Section~\ref{zookeeper_recov}. The most important reason is  
that \name{} works for the higher-performing {\em leaderless} systems while
ZooKeeper works for leader-based systems. In addition, \name{} provides three advantages.

First, \name{} provides {\em Proactive} recovery---i.e., it provides to the recovering node $R$ all the
updates that the latter missed with very low overhead. Indeed, such updates
are   collected in advance in the  \newD{}  DRAM buffer and,
on recovery, immediately sent to $R$ by its Buddy node. Second, the  \newD{}   buffer contains no redundant updates---only
the latest update to each record. This design reduces recovery overhead during
the transfer of the updates from the Buddy node to $R$,
and during their application to the persistent storage in
$R$. Finally, during recovery, $R$
remains active, participating in the protocol of ongoing client updates as a 
shadow replica, which speeds-up execution.
 
\section{Concurrent Failures}\label{sec:concurrent-fails}
As described so far, \name{} handles single failures; it is assumed that
a second failure will not occur 
until after the system has recovered from the first one. Here, we 
outline how \name{} can be extended to support concurrent failures. 

\begin{figure}[t]
    \centering
    \includegraphics[width=0.45\textwidth]{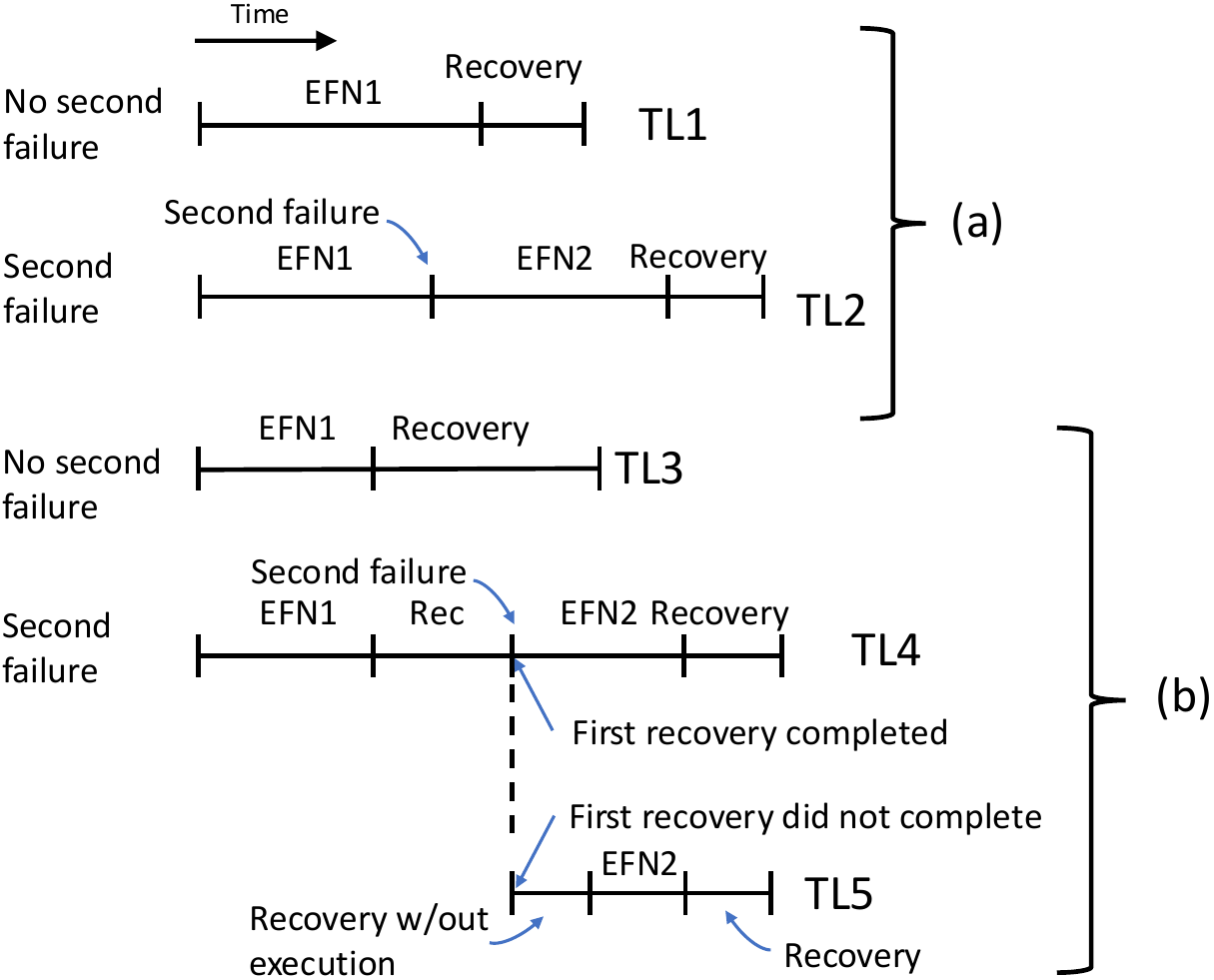} 
    \caption{Handling multiple concurrent failures. In the figure, TLi means timeline $i$.}
    \label{fig:conc-fail-timeline}
    \vspace{-6mm}
\end{figure}

Given the timeline of a node failure shown in Figure~\ref{fig:recov-timeline}, we consider
two cases: (i) during the EFN time, another node fails, and (ii) during the
Recovery time, either the Buddy node, another live node,  or the recovering node $R$ fails.
If a node fails during the EFN time, the system detects it like in the first failure. 
Because no update can make progress with a failed node, the CM reconfigures the system
again, this time entering an EFN period with two missing nodes. 
This is shown as EFN2 in Timeline TL2 in Figure~\ref{fig:conc-fail-timeline}a.

During EFN2, the live nodes continue
execution, augmenting their \newD{} buffer (which contains the updates performed
during the first, interrupted EFN1 period) with the new updates performed during this new EFN2 period.
For simplicity, the CM will wait until it can get  two working nodes to replace the two failed nodes. 
At that point, it will reconfigure the system, and
the recovery re-starts,
with two recovering nodes. Each of them picks a different 
Buddy node 
and, as it gets the \newD{} from its Buddy node, it only applies those updates that
it has missed: the node that had failed earlier applies more updates than the one that
failed later. This is shown in Figure~\ref{fig:conc-fail-timeline}a.

If a node fails during a recovery period, irrespective of whether it is a Buddy node,
$R$, or any other node, no update can make progress until
the CM  reconfigures the system. Before the second failure and reconfiguration,
$R$ may have made some progress in the recovery process. 
After the reconfiguration, \name{} checks whether $R$  is up and 
has been able to apply locally
all the updates missed since the first fault. 

If this is true (Timeline TL4 of Figure~\ref{fig:conc-fail-timeline}b),
\name{} considers the first recovery completed. 
Then, \name{} initiates an EFN2 
period while a new node is identified. When the
new node is identified and connected, a recovery  for that node is performed.   

If the first recovery did not complete (Timeline TL5 of Figure~\ref{fig:conc-fail-timeline}b), 
\name{} requires the first recovery to be redone.
For simplicity, \name{} repeats the first recovery without allowing the live nodes to continue
execution. This is reasonable given that having two concurrent failures is relatively
rare. In the special case where the second failed node is precisely 
the node that was trying to recover,
\name{} waits until a new node can be hot-plugged, 
and then proceeds to redo the first recovery. Further, in the
special case where the second failed node is precisely the Buddy node, a new
Buddy node is identified.

Once the first recovery is completed, execution resumes during an EFN2 
period while a new node is identified to fill in for the second failed node. When the
new node is identified and connected, a recovery  for the second
failure is performed.

We could instead try to redo the first recovery while the 
healthy nodes continue executing under an EFN2 period. However,
this would potentially require having a double set of \newD{} buffers per node. 

As indicated in Section~\ref{sec_fault},
if the CM fails, the Zookeeper relaunches a new instance of the CM.

\section{Protocol Verification}
\label{tlaplus-ref}

To ensure the correctness of the \hbox{\name{}} protocols, we employ the TLA+ formal specification and verify our models using the TLC model checker~\hbox{\cite{tlaplus-02}}. 
Our implementation builds on the publicly available TLA+ specification of Hermes~\hbox{\cite{hermes-asplos20}}, extending it to incorporate all $<$consistency, persistency$>$ model pairs discussed in this work.

As summarized in Table~\hbox{\ref{tab:tlaplus-properties}}, our verification covers three categories of checks. 
First, we ensure \textit{concurrency safety} by confirming the absence of deadlocks and livelocks across all reachable states. 
Second, we verify \textit{consistency correctness} by checking that the timestamp of each record in volatile memory is consistent across replicas. 
Third, we check \textit{persistency correctness} by ensuring that the timestamp of each record in NVM is the same across replicas. 
All correctness conditions are validated under the fault scenarios of Section~\hbox{\ref{sec:overview:fail-model}}, including node crashes and message loss, ensuring full system recovery.

We executed TLC on a 16-core workstation with 64\,GB of memory using breadth-first exploration and full symmetry reduction \cite{tlaplus-02}. 
For instance, the $<$Lin,~Synch$>$ verification completed in 26\,minutes, exploring 138\,million distinct states with a state-space diameter of~41. 
These values are representative of the other persistency models.

\begin{table}[t]
    \begin{scriptsize}
        \begin{tabular}{|l|}
            \hline
            \rowcolor[HTML]{EFEFEF} 
            \textbf{1. Concurrency Checks} \\ \hline
            Absence of deadlocks and livelocks. 
            \\ \hline
            \rowcolor[HTML]{EFEFEF}
            \textbf{2. Consistency Checks} \\ \hline
            a) The timestamp of the
            record in  
            volatile memory is the same across all replicas. \\ \hline
            b) When the Coordinator is about to send a VAL for consistency for a  record, the \\ 
             timestamp of the
            record in volatile
           memory is the same across all replicas. \\ \hline
            \rowcolor[HTML]{EFEFEF} 
            \textbf{3. Persistency Checks} \\ \hline
            a) The timestamp of the
            record in NVM
            is the same across all replicas. \\ \hline
            b) When the Coordinator is about to send a VAL for persistency for a record, the \\ 
            timestamp of the
            record in NVM
            is the same across all replicas. \\\hline
        \end{tabular}
       \vspace{-2mm}
        \caption{Conditions checked using TLA+ for all the models, in the presence of failures.}
        \label{tab:tlaplus-properties}
    \end{scriptsize}
    \vspace{-6mm}
\end{table}
\section{Methodology}\label{sec:methodology}\label{sec:method:workloads}

We evaluate \name{} running on a 6-node cluster with 7 cores per node
provided by CloudLab~\cite{cloudlab}. This is the largest cluster we could 
reliably obtain with the necessary configuration. The cluster parameters are shown in 
Table~\ref{tab:cloudlab-params}. One of the nodes is the CM, while the other
5 nodes run the workloads. \name{} replicates every record in each of the 
5 nodes. This is the default replication degree in past work~\cite{hermes-asplos20, minos-hpca24}. In larger machines, one could have multiple 5-node replication groups. 
The nodes in the cluster do not have modern persistent memories. Hence, we emulate the latency 
and bandwidth  
of persistent memory based on prior work~\cite{characterizing-pmem-micro20}. 
We use an in-memory datastore of 13 GB per node and model one with 
256 GB per node as will be described in Section~\ref{tao_rec}.

When a node receives a protocol message (e.g., INV), it needs to perform a sequence of steps. 
If we used the popular one-sided 
RDMA 
for communication, 
we would need multiple messages to perform these multiple steps.  Consequently, we
instead use a  Remote Procedure Call (RPC) library  named eRPC~\cite{erpc-nsdi19, erpc-code} that
provides performance similar to RDMA but with fewer transfers. 
The CM uses one-sided RDMA for the heartbeat exchange. 

\begin{table}[t]
    \centering
    \caption{Cluster used to run \name{}.}
    \vspace*{-.6\baselineskip}
    \begin{scriptsize}
    \begin{tabular}{|l|l|}
        \hline
        Number of nodes & 6 \\ \hline
        CPU per node & Xeon E5-2450 (7 cores, 2.1 GHz) \\ \hline
        In-memory datastore & 13GB/node and 256GB/node (modeled)\\  \hline
        NIC per node & Mellanox MX354A FDR CX3 \\ \hline
        Timeouts & 2 ms INV-ACK, 2 ms ACK-VAL\\ \hline
        Heartbeat period & 4 ms \\ \hline
    \end{tabular}
    \end{scriptsize}
    \label{tab:cloudlab-params}
    \vspace*{-.4\baselineskip}
    \vspace{-4mm}
\end{table}

\noindent
\textbf{Workloads Used.}
We evaluate \name{} with two different benchmarks: 
the TAOBench application suite~\cite{taobench-vldb22} and
the Yahoo! Cloud Serving Benchmark (YCSB)~\cite{ycsb-socc10,ycsb-code-2014}.
 TAOBench  tests the performance of social graph databases, simulating real-world user interactions. 
In our experiments, we use four of the provided drivers: MySQL~\cite{mysql}, CockroachDB (CRDB)~\cite{crdb-sigmod20}, YugabyteDB~\cite{yugabyte-2022}, and Cloud Spanner (Spanner)~\cite{spanner-2022}. 
The default setup uses a uniform record distribution, and 80\% reads and 20\% writes.
After failure detection, we set the period of
execution with fewer nodes (EFN) to  360 seconds.

Because the TAOBench runs take a long time, we also run YCSB.
For the YCSB experiments, we use a Hashtable~\cite{bigtable-acm08} back-end in-memory 
application. 
The workload uses a zipfian distribution for records, has 
80\% read and
20\% write  operations, and each node issues
3M back-to-back operations. 
Because YCSB are synthetic benchmarks that issue back-to-back transactions,
we set their EFN period to 6 seconds.

\noindent
\textbf{Systems Evaluated.}
We evaluate three leaderless systems that use the Hermes~\cite{hermes-asplos20} consistency transactions
on a MINOS\cite{minos-hpca24} implementation of different persistency models:
{\em Baseline}, \textit{\name}, and \textit{State-of-the-art}.
{\em Baseline} does not have any recovery support and we assume that
it suffers no failure.
\textit{\name{}} is Baseline plus our proposed
incremental recovery mechanism.
\textit{State-of-the-art} is Baseline with a nonincremental recovery;
on a fault, it copies the full database from one node to another like Hermes. We cannot use Hermes as is because Hermes
does not consider any persistency model.
In both \name{} and State-of-the-art, we model a node failure and subsequent
system recovery. The default DDP model used is $<$Lin, Synch$>$.
We are interested in two metrics: 1) failure recovery latency and 2) loss of system throughput due to the 
failure.

In Section~\ref{compare_zoo}, we also compare \name{} to the ZooKeeper online incremental recovery scheme. 
The two systems differ not only on the three \name{} recovery optimizations
described in Section~\ref{zookeeper_compare}; they also differ in the protocol
used: \name{} is {\em leaderless} and ZooKeeper is {\em leader-based}.
\section{Evaluation}\label{sec:evaluation}

We first 
use TAOBench to compare {\em \name}'s recovery latency and throughput degradation
  to  {\em State-of-the-art}.
 Then, we use the
faster YCSB runs to characterize \name. Finally, we compare \name{} to the
leader-based ZooKeeper.

\subsection{Recovery Latency with the TAOBench Suite}
\label{tao_rec}

\begin{figure}[t]
    \centering
    \includegraphics[width=0.4\textwidth]{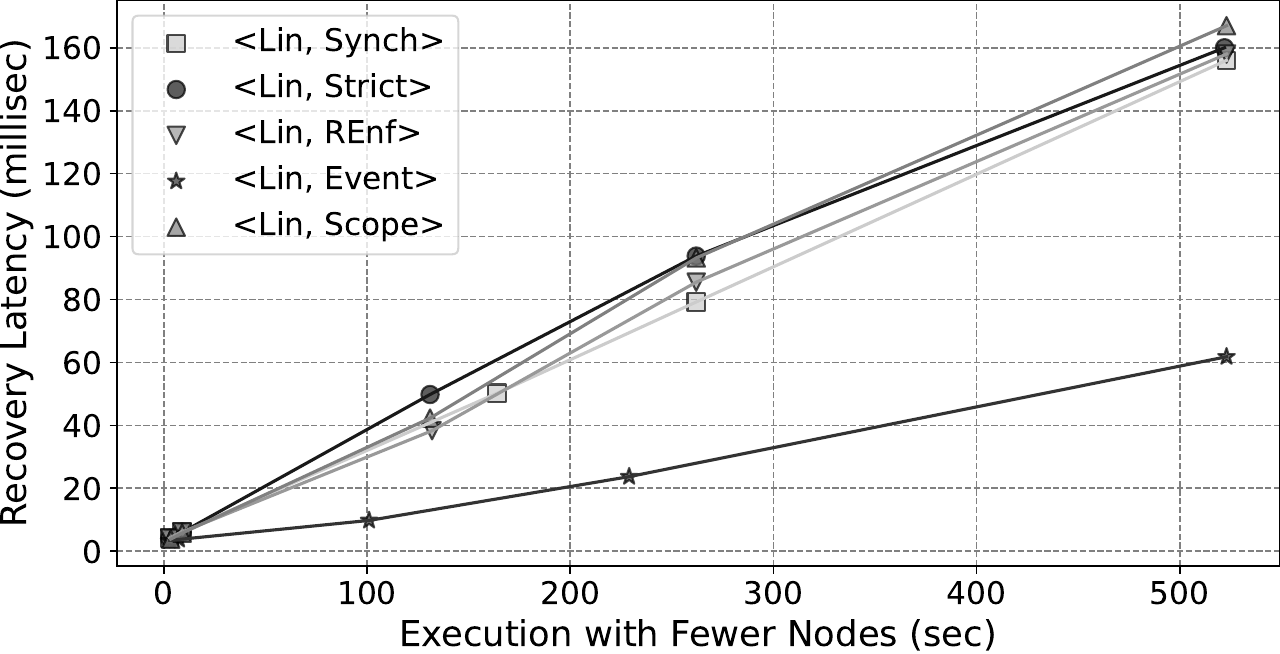} 
     \vspace{-2mm}
    \caption{\name{} recovery latency as a function of the EFN duration for a 256GB datastore per node.}
    \label{fig:eval:recov-lat-efn256GB}
    \vspace{-4mm}
\end{figure}

Figure~\ref{fig:eval:recov-lat-efn256GB} shows 
the recovery latency of \name{} as a function of the duration of the execution with fewer nodes (EFN).
The figure refers to the average of our four TAOBench drivers, and shows   curves for the 
different persistency models. We model a 256GB datastore per node.
Since the machine we use for our experiments only has 16GB per node, we cannot execute the actual
datastore. Hence, we use  a 13GB datastore per node but emulate a very
large datastore by assuming
that all the updates performed during the EFN period are to {\em different keys}. Hence, the 
\newD{} buffer adds an entry at every update, and its size grows fast, which increases the 
recovery time. 

We see that, as the EFN period increases, the recovery that
follows it takes a longer time 
in all cases. This is because, the longer a node is offline, the more state 
the \newD{} buffers accumulate; then,
during recovery, this state is applied to the recovering node. We also see that the recovery 
latency is similar for all   models except  $<$Lin, Event$>$. The latter has
a faster recovery mostly because, in Event, the recovering node 
applies the updates received from its Buddy node to its  persistent memory in the background.

\begin{figure*}[t]
    \centering
    \includegraphics[width=.92\textwidth]{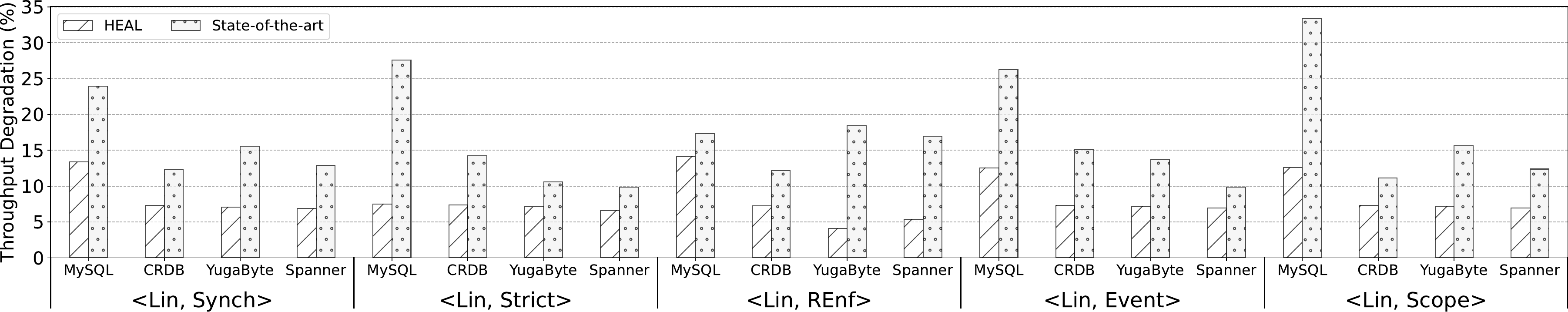}
    \vspace{-2mm}
    \caption{Throughput degradation of {\em \name}  and {\em State-of-the-art} over {\em Baseline}  running TAOBench for a 256 GB datastore. Lower values are better.} 
    \label{fig:eval:tao-bench}
   \vspace*{-.5\baselineskip}
    \vspace{-3mm}
\end{figure*}

We see that the recovery latency is always small.
For example, for  an EFN period of
360s, the recovery latency of $<$Lin, Synch$>$ is about 120ms. This is in contrast to the 
time it takes for {\em State-of-the-art} to transfer the database from
a recoverer node’s persistent memory (PM) to the recovering node’s PM. From
Figure~\ref{fig:hermes-recov}, such time is 360 sec  for 
DRAM+PM and 254 
sec  for PM-only in a 256GB datastore per node. These times are
lower  bounds of the recovery latency of {\em State-of-the-art}.

Table~\ref{tab:missedUpdSz} shows the size of the per-node \newD{} buffers for
 Lin consistency and various persistency models for experiments with
 13GB and 256GB datastores.
The data is the average across all the TAOBench drivers.
We see that, for a 13GB database, the \newD{} buffers are  small. 
This is because the datastore is   small, and many keys are reused during the EFN period. 
In contrast, for a 256GB datastore, the \newD{} buffer sizes are larger,
and reach 17.7--32.1 MB per node. In addition, the sizes vary across the models, which
we attribute to the fact that the EFN period measured in different experiments 
corresponds to
somewhat different sections of the program. Overall, this is an acceptable
cost to enable incremental recovery. If the EFN period was so long that the \newD{} buffer overflowed,
the system would revert  to recovering by transferring the whole datastore.

\begin{table}[t]
    
    \centering
    \caption{Average per-node \newD{} buffer size for Lin consistency 
    and various persistency models.}
    \vspace*{-.6\baselineskip}
    \begin{scriptsize}
    \begin{tabular}{|l|ccccc|}
    \hline
    \rowcolor[HTML]{EFEFEF} 
    \cellcolor[HTML]{EFEFEF} & \multicolumn{5}{c|}{\cellcolor[HTML]{EFEFEF}\textbf{Persistency Model}} \\ \hline
    \rowcolor[HTML]{EFEFEF} 
    \multirow{-2}{*}{\cellcolor[HTML]{EFEFEF}\textbf{\begin{tabular}[c]{@{}l@{}}Datastore\\ Size\end{tabular}}} & \multicolumn{1}{c|}{\cellcolor[HTML]{EFEFEF}\textbf{Synch}} & \multicolumn{1}{c|}{\cellcolor[HTML]{EFEFEF}\textbf{Strict}} & \multicolumn{1}{c|}{\cellcolor[HTML]{EFEFEF}\textbf{REnf}} & \multicolumn{1}{c|}{\cellcolor[HTML]{EFEFEF}\textbf{Event}} & \textbf{Scope} \\ \hline
    \textbf{13GB} & \multicolumn{1}{c|}{2.25 MB} & \multicolumn{1}{c|}{2.25 MB} & \multicolumn{1}{c|}{2.25 MB} & \multicolumn{1}{c|}{2.25 MB} & 2.26 MB \\ \hline
    \textbf{256GB} & \multicolumn{1}{c|}{19.20 MB} & \multicolumn{1}{c|}{28.86 MB} & \multicolumn{1}{c|}{17.66 MB} & \multicolumn{1}{c|}{18.11 MB} & 32.14 MB \\ \hline
    \end{tabular}
    \end{scriptsize}
    \label{tab:missedUpdSz}
     \vspace{-5mm}
\end{table}

\vspace{-1mm}
\subsection{Throughput with the TAOBench Suite}
\label{sec:eval:tao}

Figure~\ref{fig:eval:tao-bench} shows the throughput degradation of {\em \name} and {\em State-of-the-art}
over {\em Baseline} for  the 4 TAOBench drivers  and 5 models.  {\em Baseline} has no 
failure, while {\em \name} and {\em State-of-the-art} suffer and recover from one node failure. 
The {\em State-of-the-art} bars use the {\em DRAM+PM} 
environment of Fig.~\ref{fig:hermes-recov}.
We model the
256GB datastore. In the figure,
lower bars are better.

If we consider $<$Lin, Synch$>$, we see that the average throughput degradation in 
{\em \name} and {\em State-of-the-art} is 8.7\% and 16.2\%, respectively. 
Across all persistency models, the throughput degradation in 
{\em \name} is 4--14\%, with an average
of 8.1\%; the throughput degradation in {\em State-of-the-art}
is 10--34\%, with an average of 16.5\%. 
The reason for the difference 
between {\em \name} and {\em State-of-the-art} is the different 
recovery latencies in the two systems. In both systems, the application starts
with  healthy nodes, suffers the loss of a node for 360s, goes through a recovery, and then
finishes execution with all nodes healthy. During the EFN and recovery periods, both systems 
execute with lower throughput. Since the recovery completes 
much faster in {\em \name}, the average throughput degradation in  
{\em \name} is substantially
 lower than in {\em State-of-the-art}.

The difference in throughput degradation between {\em \name} and {\em State-of-the-art}
remains large across persistency models.
In general, weaker models such as $<$Lin, Event$>$ should suffer less
absolute throughput degradation since, as seen in Sec.~\ref{tao_rec}, their recovery
latency is shorter. However, weaker models also   execute faster in {\em Baseline}, and
the loss in throughput during the constant-time EFN period causes a higher loss in 
weaker models. For these reasons, the percentage throughput degradation is similar
across persistency models.
 
\begin{figure*}[t]
    \centering
     \includegraphics[width=.65\textwidth]{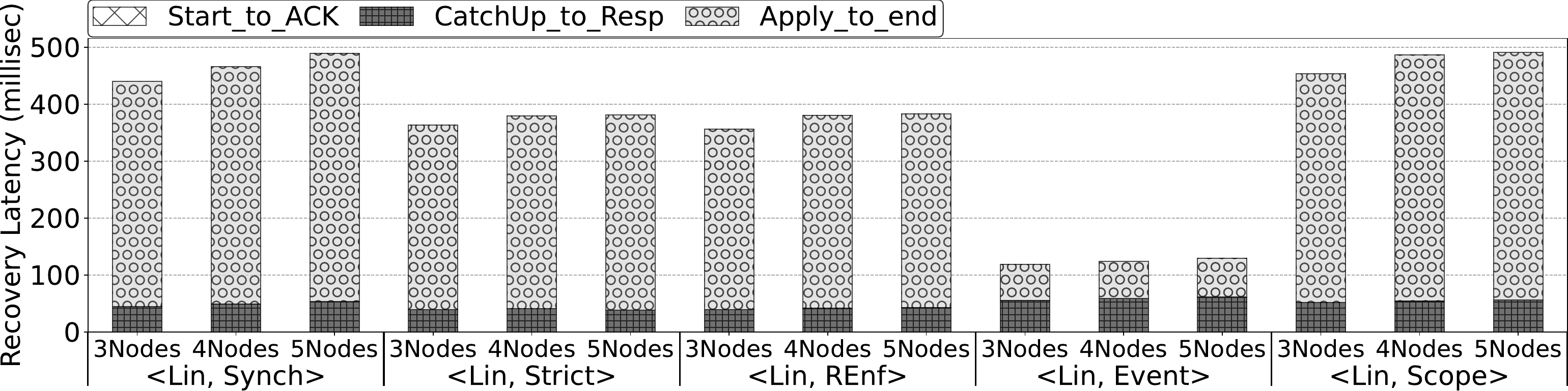} 
    \caption{Breakdown of {\em \name}'s  recovery latency   for different 
    persistency models and numbers of nodes.}
    \label{fig:eval:recov-bkdown}
    \vspace{-5mm}
\end{figure*}

\begin{figure}[t]
    \centering
    \begin{subfigure}[b]{0.4\textwidth}
        \includegraphics[width=.9\textwidth]{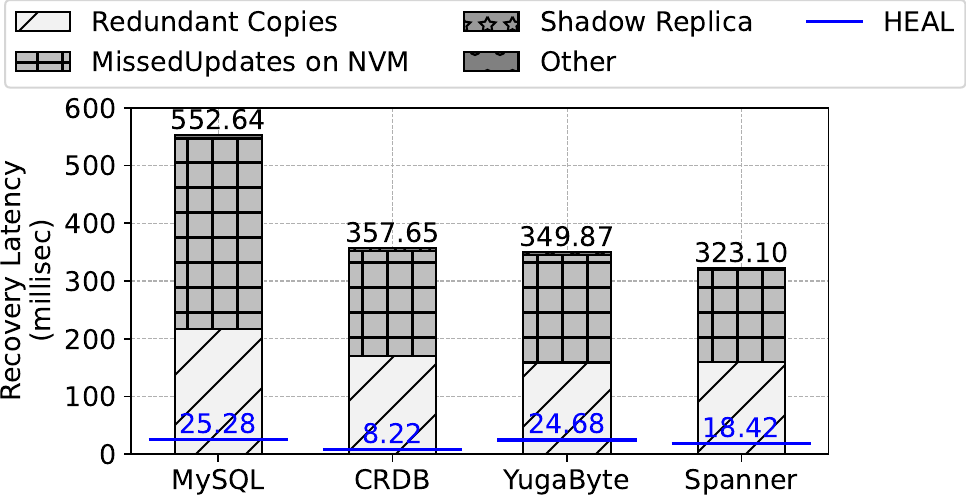} 
        \label{fig:zk-subfig-a}
    \end{subfigure}
    \begin{subfigure}[b]{0.4\textwidth}
        \includegraphics[width=.78\textwidth]{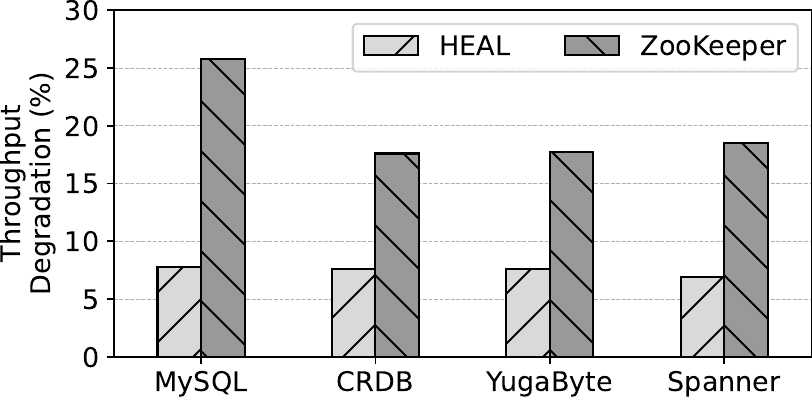}
        \label{fig:zk-subfig-b}
    \end{subfigure}
    \vspace{-1mm}
    \caption{{\em \name} vs ZooKeeper for a 13 GB datastore: recovery latency (top) and throughput degradation over the failure-free {\em Baseline} (bottom).}
    \label{fig:zookeeper}
 \vspace{-7mm}
\end{figure}

\vspace{-1mm}
\subsection{Breakdown of the \name{} Recovery Latency}

We now characterize {\em \name} with faster runs using  YCSB. 
Figure~\ref{fig:eval:recov-bkdown} shows the recovery latency 
for all models and for executions with 5, 4, or 3 nodes.
The bars are broken down into the three steps of the
recovery shown in Fig.~\ref{fig:recov-recovering-node}: from the beginning to
when the recovering node $R$ receives all the Recov\_start\_ACKs 
({\em Start\_to\_ACK}), from when $R$ prepares and sends 
CatchUp\_req  until it receives   CatchUp\_resp  
({\em CatchUp\_to\_Resp}),
and from when $R$ starts applying the updates based on
CatchUp\_resp until the end ({\em Apply\_to\_end}). 

The Start\_to\_ACK time is negligible, as it involves exchanging small messages.  
CatchUp\_to\_Resp  takes about 10-15\% of the recovery latency in most models
and about 50\% in $<$Lin, Event$>$. In absolute terms, however,  CatchUp\_to\_Resp  
 is similar in all models, as it involves sending at least one  bulky message.
Apply\_to\_end  is typically the  dominant time by far. The only exception
is for the weak $<$Lin, Event$>$. Apply\_to\_end is long because the recovering
node $R$ persists  locally all the entries in the \newD{} buffer that are not obsolete.
In $<$Lin, Event$>$,  Apply\_to\_end is short because, under Event persistency, 
 $R$ does not apply the updates to the persistent memory
in the critical path. 

$<$Lin, Scope$>$ has a high  Apply\_to\_end time even though it is  
weaker than some other models. The reason may be because this
model needs more synchronization between nodes through   [PERSIST]sc commands.
The Apply\_to\_end time tends to increase with the node count
due to the higher network activity.

\subsection{Comparison to the Leader-based ZooKeeper Scheme}
\label{compare_zoo}
 
Fig.~\ref{fig:zookeeper} compares {\em \name} with $<$Lin, Synch$>$ and ZooKeeper.
We run TAOBench on
a 13 GB datastore, as the system 
suffers a node failure and, after a 360-second EFN period, recovers. The top plot
shows the recovery latency of ZooKeeper (in bars)  and {\em \name} (as horizontal blue lines). 
 The ZooKeeper latency is broken into four components. The first three correspond to the shortcomings in ZooKeeper that   {\em \name} eliminates (Sections~\ref{zookeeper_recov} and~\ref{zookeeper_compare}),
 slightly modified for ease of measurement: 
 1) redundant updates in the leader's log that are read from non-volatile storage, transferred, and applied to the 
 recovering node's log ({\em Redundant Copies}),
 2) accessing the leader's log in non-volatile storage rather than  a proactively-prepared
 DRAM buffer {\em for the non-redundant copies} 
({\em MissedUpdates on NVM}),
and not allowing the recovering node to transition  to a shadow replica ({\em Shadow Replica}).
The fourth category is other recovery operations ({\em Other}). 

We see that {\em \name} recovers substantially faster than the leader-based
ZooKeeper scheme. On average, its recovery latency is 20.7$\times$ 
lower. Consider now the three identified overheads.
Manipulating redundant copies in time-consuming non-volatile storage accesses and data transfers is a major overhead in ZooKeeper. Maintaining the log in non-volatile storage for the  remaining, non redundant copies  is also a large overhead. However,
the absence of shadow replica operations incurs minimal additional overhead.
In part, this is because these workloads are read-intensive; in 
write-intensive ones, the overhead will be higher. Finally, the
{\em Other} category  is 
primarily associated with ZooKeeper's leader-based architecture. It
accounts for about 40\% of the recovery latency.

The bottom plot shows the  
throughput degradation experienced by applications over the {\em Baseline}.
On average, \name{} reduces the throughput degradation by 62.4\% over ZooKeeper. While both schemes
use on-line incremental recovery, their main 
throughput differences stem from:
(i) \name{} being leaderless and 
ZooKeeper being leader-based, and (ii) the two main inefficiencies
of the ZooKeeper recovery protocol described above that 
can be eliminated with some \name{} techniques.
\section{Related Work}\label{sec:relatedwork}

Many recovery techniques 
have been proposed. As indicated before, they use
logging~\cite{singlenode-osdi14, adaptivelogging-sigmod16, 
pacman-sigmod17, skyros-sosp21}, replication~\cite{hermes-asplos20, epaxos-nsdi21}, erasure coding~\cite{paritycheck-fast14},
or a combination of them.
Most of the current systems operate in leader-based mode (e.g.,~\cite{zookeeper-atc10,farm-nsdi14,ramcloud-sosp11,skyros-sosp21,cad-fast20}). They use different consistency and persistency 
models---e.g., RAMCloud~\cite{ramcloud-sosp11} and Skyros~\cite{skyros-sosp21} support a model similar to   $<$Lin, Synch$>$; CAD~\cite{cad-fast20} supports a persistency model like REnf.

Several transactional~\cite{dsn01, srds02} and State Machine Replication  (SMR)~\cite{icdcs17} 
schemes rely on clients broadcasting each transaction or operation
to all the nodes in the cluster, and ordering transactions/operations through a Global
ID. For that, they use special software support for a broadcast primitive, usually through consensus~\cite{icdcs17}. Such support is expensive.

There are only a few systems that operate in the recent, high-performance 
leaderless mode. They include EPaxos~\cite{epaxos-nsdi21}, Hermes~\cite{hermes-asplos20}, and MINOS~\cite{minos-hpca24}.
These schemes attain the highest-performance execution, but  make online incremental recovery more complex.
In Hermes, recovery after a node failure
involves a full database transfer to the node being recovered.
Moreover, Hermes does not consider the concept of 
durability. EPaxos describes a leaderless consensus algorithm, but does not consider recovery.
\name{} is  
the first online, incremental recovery scheme for leaderless
 systems. Its protocol is based on MINOS~\cite{minos-hpca24} and is 
optimized for minimal recovery latency and low throughput impact. 

\section{Discussion}
\label{sec::discussion}
\name{} builds on prior work that assumes byte-addressable NVM as the durable layer \cite{minos-hpca24}.
Using  SSDs instead of NVM is orthogonal to \name{}'s protocol. It 
increases  persistency delays due to file I/O, but does not affect  
correctness or   semantics.

Empirical studies show that node failures rarely corrupt the local durable state. 
Past literature assumes 
that, in node outages,   
the local persistent 
storage remains  intact~\cite{saucr-osdi18, zhuque-atc23}. 
This observation motivates \name{}'s assumption that durable state can be reused after recovery.

\section{Conclusion}\label{sec:conclusion}
 
This paper proposed {\em \name{}}, 
the first online, incremental recovery scheme  for 
 leaderless distributed systems. It introduced
algorithms for linearizable consistency and 5 persistency models. 
On average,  \name{}  
recovers a cluster after 
a node failure in
120 ms, while reducing the throughput 
by 8.7\%. 
In comparison, a  state-of-the-art 
recovery scheme for leaderless
systems needs 360 seconds
to recover, and reduces the throughput 
by 16.2\%.
Further, compared to a leader-based incremental recovery scheme,
\name{} reduces the 
recovery 
latency by 20.7$\times$
and the throughput degradation by 62.4\%.



\bibliographystyle{IEEEtranS}
\bibliography{refs}

\end{document}